\documentclass[12pt]{article}
\newcommand{\lb}[1]{\label{#1}}
\newcommand{\be}{\begin{equation}}
\newcommand{\ee}{\end{equation}}
\newcommand{\ba}[1]{\begin{array}{#1}}
\newcommand{\ea}{\end{array}}

\usepackage{graphicx}
\usepackage{amssymb}
\usepackage{amsfonts}
\usepackage{epsfig}

\begin{document}

\thispagestyle{empty}

\begin{flushright}

{\small {\tt cond-mat/0011396v2}\\{\sl December 2005}}

\end{flushright}

\begin{center}

{\large\bf{Fractal Structure of the Harper Map Phase Diagram from\\ Topological Hierarchical Classification}}\\[10mm]

\vspace{1 truecm}

P. Castelo Ferreira\footnote{Adrss. for original paper: Dept. of Physics, Th. Phys., University of Oxford, Oxford OX1 3NP, U.K.\\ Curr. Adrss.: Dept. de F\'{\i}s., Univ. da Beira Interior, Av. Marqu\^es D'\'Avila e Bolama, 6201-001 Covilh\~a\\ Curr. Adrss.: CENTRA, IST, Av. Rovisco Pais, 1049-149 Lisboa, Portugal}\\[25mm]

{\bf\sc Abstract}
\begin{minipage}{15cm}
\vspace{2mm}
It is suggested a topological hierarchical classification
of the infinite many Localized phases figuring in the phase diagram
of the Harper equation for anisotropy parameter $\epsilon$ versus Energy $E$ with
irrational magnetic flux $\omega$. It is also proposed a rule that explain
the fractal structure of the phase diagram.
Among many other applications, this system is equivalent to the
Semi-classical problem of Bloch electrons in a
uniform magnetic field, the Azbel-Hofstadter model, where the discrete magnetic translations operators
constitute the quantum algebra $U_q(sl_2)$ with $q^2=e^{i2\pi\omega}$.
The magnetic flux is taken to be the golden mean
$\omega^*=(\sqrt{5}-1)/2$ and is obtained by successive rational
approximants $\omega_m=F_{m-1}/F_m$ with $F_m$ given by the Fibonacci sequence $F_m$.[OUTP-00-08S, \texttt{cond-mat/0011396}]
\end{minipage}

\end{center}

\noindent

\vfill
\begin{flushleft}
\small{Keywords: Topological Classification, Harper, Azbel-Hofstadter, Phase Diagram, Fractal, Quantum Groups}
\end{flushleft}

\newpage

\section{Introduction}
In this article we study the phase diagram $E\times\epsilon$ of the Harper equation~\cite{Harper55} 
\be
\ba{rcl}
\psi_{n+1}+\psi_{n-1}+2\epsilon\cos(2\pi\phi_n)\psi_n&=&E\psi_n\ ,\\[5mm]
\phi_n&=&\phi_0+\omega n\ .
\ea
\lb{schro}
\ee
The Harper equation is a Schr\"odinger operator of two discrete variables $\psi$ and $\phi$ and, among other
application, is associated with the quantum algebra of discrete magnetic translations $U_q(sl_2)$. Namely it is
obtained in the Azbel-Hofstadter model~\cite{AZ_0,HO_0} of Bloch electrons
(electrons on a $2$-dimensional periodic potential) in the presence of a
uniform magnetic field in the Landau gauge. Here the anisotropy parameter $\epsilon$ is the
ratio between the spacing in the two directions of the underlying $2$-dimensional periodic potential lattice,
$\omega$ is the magnetic flux per unit cell of the potential lattice
and the quantum deformation parameter is $q^2=e^{i2\pi\omega}$~\cite{AZ_0,WZ_0}.

The Harper equation is second order on $\psi$. In order to simplify the numerical treatment
of this equation we can cast it into a first order equation
by considering the transformation of variables $x_n=\psi_{n-1}/\psi_n$. The resulting equation is equivalent
to the Harper equation~(\ref{schro}) and is known as the Harper map~\cite{KS_1}
\be
\ba{rcl}
x_{n+1}&=&\displaystyle\frac{-1}{x_n-E+2\epsilon\cos(2\pi\phi_n)}\ ,\\[5mm]
\phi_n&=&\phi_0+\omega n\ .
\ea
\lb{harper}
\ee

In order to study the phase diagram $E\times\epsilon$ we need to identify the Localized and Extended phases. In order to do so
we use Aubry and Andr\'e study~\cite{aa}, they proved that the Lyapunov exponent $\lambda$ is proportional to the localization
length, specifically $\lambda=-2\gamma$. For the Harper map $\lambda\le 0$, such that for $\lambda=0$ the phase 
is Extended and for $\lambda<0$ the phase is Localized. The Lyapunov exponent corresponding
to the $\phi$ dynamics is always $0$ and the one concerning to the $x$ dynamics is given by
\be
\lambda=\lim_{N\to\infty}\frac{1}{N}\sum^{N}_{n=0}y_n\ ,
\ee
where $y_n=\log\frac{\partial x_{n+1}}{\partial x_n}=\log x_{n+1}^2$.

There exists already in the literature classifications, both for the Extended phases~\cite{WZ_1} and
for the Localized phases~\cite{ours} of equations~(\ref{schro}) and~(\ref{harper}).
In the remaining of this article we propose a new hierarchical topological classification
of the several Localized phases for irrational flux $\omega^*=(\sqrt{5}-1)/2$ based on the
topological winding number of the attractors on the space of variables $(\phi,x)$. In order to study
the irrational phase diagram we take successive rational approximants $\omega_m$.
The fractal structure of the phase diagram emerge naturally in this
framework and we suggest a rule that reproduces it. As we will show in detail, the great advantage
of this new classification is that its labels correspond to the magnitude of the several different
phases. Here the phase magnitude is understood as the area of each different phase on the full phase
diagram. Also the area (hence magnitude) of the several phases coincide with its order of appearance
in the phase diagrams for successive approximants $\omega_m$ to $\omega^*$.

This article is organized as follows, in section~2 we introduce the irrational phase diagram
and successive approximants as well as the one-dimensional attractor in the $\phi\times x$
diagram for irrational $\omega^*$. In section~3 we justify the identification of the $\phi\times x$
with a torus and introduce our classification. In section~4 we analyse the phase diagram and
give the rules to build the Hierarchical Tree for the phase diagram.
In order to better explain the fractal structure and how to build this Tree
we have three appendixes. In appendix~\ref{A.A} we
present the Lyapunov exponents for $\omega=\omega^*$ and $\epsilon=1$
with enough resolution to identify the several phases (that correspond to each negative \textit{bump}).
In appendix~{\ref{A.B}} we list the $\phi\times x$ diagrams up to $\eta=55$ (with indication of the respective energies).
Finally in appendix~\ref{A.C} we show graphically how to build the triangular structure up to level 9.

\section{Phase Diagram\lb{sec:phases}}
We proceed to present the numerical results and study the properties of the $E\times\epsilon$ phase
diagram of the Harper map~(\ref{harper}) which is equivalent to the Harper equation~(\ref{schro}).
We consider successive rational approximants $\omega=\omega_m$ of 
the irrational value $\omega=\omega^*$ given by the golden mean.
The rational succession $\omega_m=F_{m-1}/F_m$ converges
to the irrational golden mean $\omega^*=(\sqrt{5}-1)/2$. Being $F_0=F_1=1$
and $F_m=F_{m-1}+F_{m-2}$ the $m^{\mathrm{th}}$ Fibonacci number in the Fibonacci sequence.

\begin{figure}[ht]
 \begin{picture}(65,40)(-10,1)
  \put(-25,-151){\includegraphics[width=280mm]{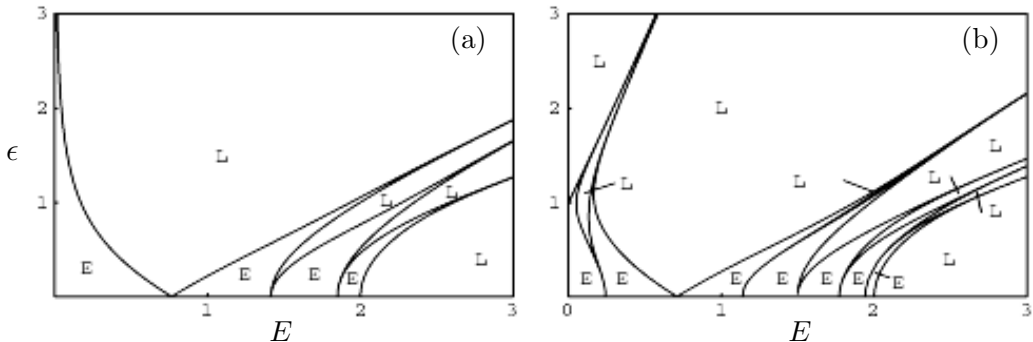}}
  \put(106,4){\small $E$}
  \put(37,4){\small $E$}
  \put(2,28){$\epsilon$}
  \put(61,43){\small (a)}
  \put(129,43){\small (b)}
 \end{picture}
\caption{\footnotesize Phase diagram for rational $\omega$ for 
$\phi_0=0$. The Localized phases (L) are presented.
(a) $\omega_5=5/8$; (b) $\omega_6=8/13$.}
\label{fig:phaser}
\end{figure}

Computing the phase diagram for successive rational $\omega_m$ allows us to
\textit{visualize} the fractal structure of the irrational $\omega^*$
emerging in the phase diagrams. As we consider successive rational approximants, the Extended phases
split and new thinner Localized phases emerge. Their number
increases until it becomes infinite. More precisely
for a given $\omega_m$ there are $F_{m-1}$ Localized phases in the first quadrant.
Two of such diagrams are pictured in figure~\ref{fig:phaser}
for $\omega=5/8$ and $\omega=8/13$.
For greater $m$'s the phase transition lines become denser
and denser. In the limit of irrational $\omega^*$, the lines cross at the critical value
$\epsilon=1$ and the Lyapunov exponents go to zero at these points
of intersection~\cite{PS_1}. Although for the rational values $\omega_m$ the phase diagrams depend on
$\phi_0$, in the irrational limit $\omega^*$ the phase diagram is not sensitive to $\phi_0$.
\begin{figure}[ht]
 \begin{picture}(73,95)(-27,10)
  \put(-6,-26){\includegraphics[width=150mm]{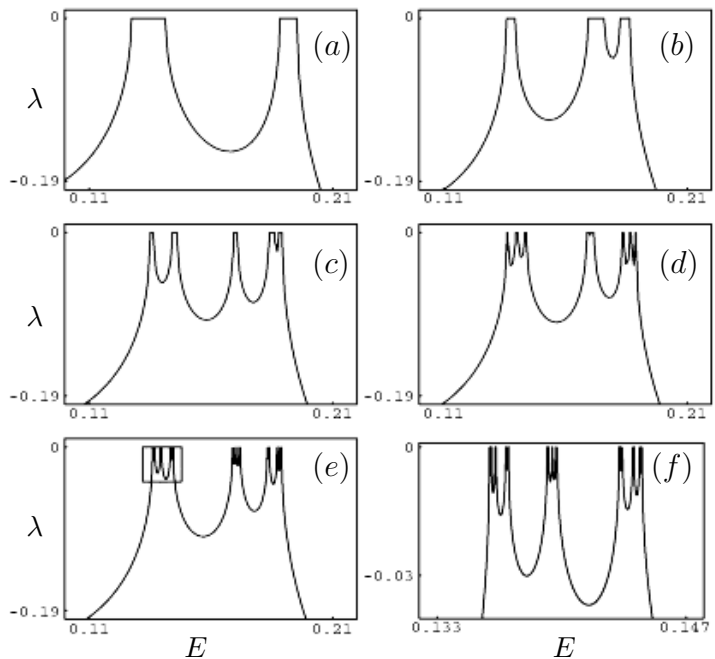}}
  \put(28,10){\small $E$}
  \put(77,10){\small $E$}
  \put(7,26){\small $\lambda$}
  \put(7,54){\small $\lambda$}
  \put(7,83){\small $\lambda$}
  \put(45,34){$(e)$}
  \put(45,61){$(c)$}
  \put(45,90){$(a)$}
  \put(90,34){$(f)$}
  \put(91,61){$(d)$}
  \put(91,90){$(b)$}
 \end{picture}
\caption{\footnotesize Fractalization of the phase diagram. 
Lyapunov exponents $\lambda$ for $\epsilon=1$ and $\phi_0=0$.
(a) $\omega_7=13/21$; (b) $\omega_8=21/34$; (c) $\omega_9=34/55$;
(d) $\omega_{10}=55/89$; (e) $\omega=\omega^*=(\sqrt{5}-1)/2$; (f) $\omega=\omega^*$, zoom of (e).}
\label{fig:lyaw}
\end{figure}
Note also that for $\epsilon>1$ the Lyapunov exponents become always
negative, this means that the phase transitions
are smooth without undergoing an Extended phase. See~\cite{ours} for
further details in these discussion.
As a final remark we note that the phase diagram is symmetric 
with respect to both axes $E=0$ and $\epsilon=0$. Also we stress that from the mathematical point of
view the parameters $E$ and $\epsilon$ can be both negative or positive, from a physical
point of view they should be positive, $E$ is the energy and $\epsilon$ is the ratio of
the unit cell sides.
\begin{figure}[ht]
 \begin{picture}(73,85)(-20,-10)
  \put(-8,-80){\includegraphics[width=220mm]{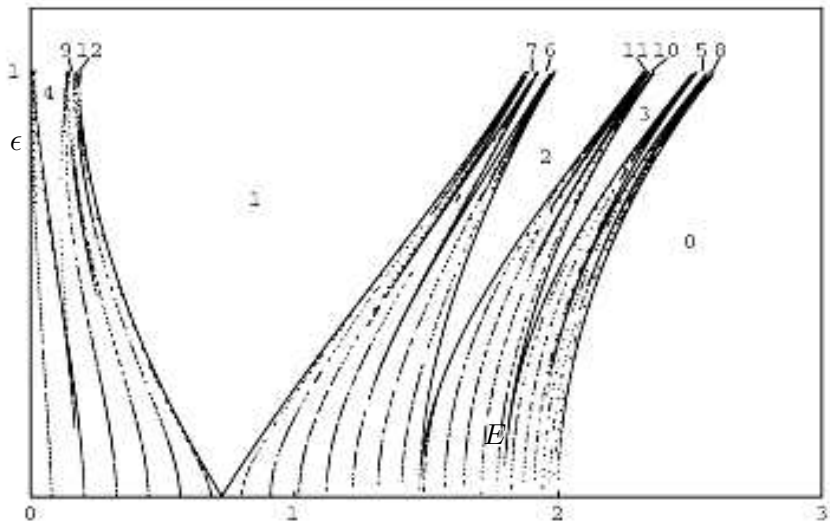}}
  \put(68,0){\small $E$}
  \put(5,39){$\epsilon$}
 \end{picture}
\caption{\footnotesize Phase diagram for $\omega^*$. The first 12
zones are labeled in the diagram.}
\label{fig:phaseir}
\end{figure}
The phase diagram becomes in the limit of irrational $\omega^*$ a fractal
(see figure~\ref{fig:lyaw} and~\ref{fig:phaseir}), more exactly, a
Cantor set~\cite{bs}. It is this particular limit that interest us. From the
above discussion it is enough to restrict our analysis of the several phases to the first
quadrant for $\epsilon<1$ and $\phi_0=0$.

It remains to study the attractors in the space of variables $(\phi,x)$.
Concerning the purpose of this work we describe
the behaviour of the map in the Localized phases only, for further details on extended phases see~\cite{WZ_1,ours}.
For a given $\omega_m$, the map has 0-dimensional attractors with period $F_m$, this means that the
attractor is a $F_m$-cycle. In other words the map iteration cycle trough $F_m$ distinct points.
In the limit $\omega\to\omega^*$ it becomes an one-dimensional attractor $F_\infty$, it is a line.
These results are present in figure~\ref{fig:attractor}.
It is this property that we are exploring next in order to define our phase classification.
\begin{figure}[ht]
 \begin{picture}(73,61)(-29,6)
  \put(-5,-29){\includegraphics[width=140mm]{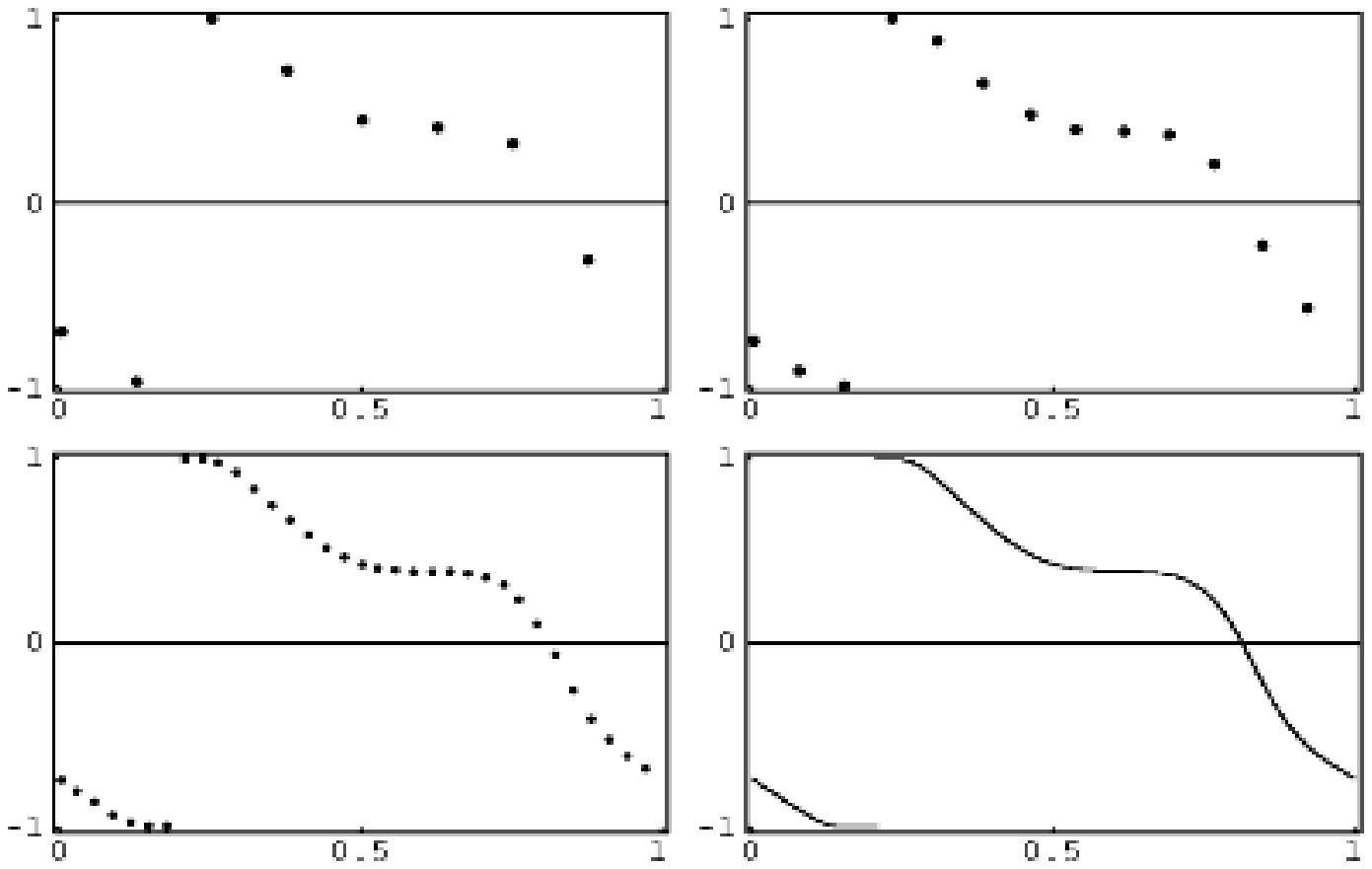}}
  \put(70,4){\small $\phi$}
  \put(0,44){\rotatebox{90}{\small $\tanh(x)$}}
  \put(27,4){\small $\phi$}
  \put(0,14){\rotatebox{90}{\small $\tanh(x)$}}
  \put(43,57){\small (a)}
  \put(86,57){\small (b)}
  \put(43,29){\small (c)}
  \put(86,29){\small (d)}
 \end{picture}
\caption{\footnotesize Attractors in the space of variables $\phi\times x$ corresponding to the
Localized phase $1$, for $\phi_0=0$, $E=0.75$ and $\epsilon=0.5$.
(a)~$\omega_5=5/8$; (b)~$\omega_6=8/13$; (c)~$\omega_{11}=89/144$;
(d)~$\omega=\omega^*=(\sqrt{5}-1)/2$, in this limit the $F_m$-cycle becomes a line.}
\lb{fig:attractor}
\end{figure}

\section{Topological Classification}
It is time to define the classification.
We are considering a map from the generic space of variables $\phi\times x$
to a torus such that the identifications
$\phi:\ 0\cong 1$ and $x:\ +\infty\cong-\infty$ hold.
We note that we don't actually have to impose these identifications,
they naturally come from the definitions of the variables $\phi$ and $x$.

The $\phi$ identification is justified by
noting that $2\pi\phi$ is an angular phase, then the variable
$\phi$ is compact in the interval $[0,1]$ with identification $\phi:\ 0\cong 1$
as any usual angular variable.
\begin{figure}[ht]
 \begin{picture}(73,55)(-10,0)
  \put(-3,-75){\includegraphics[width=175mm]{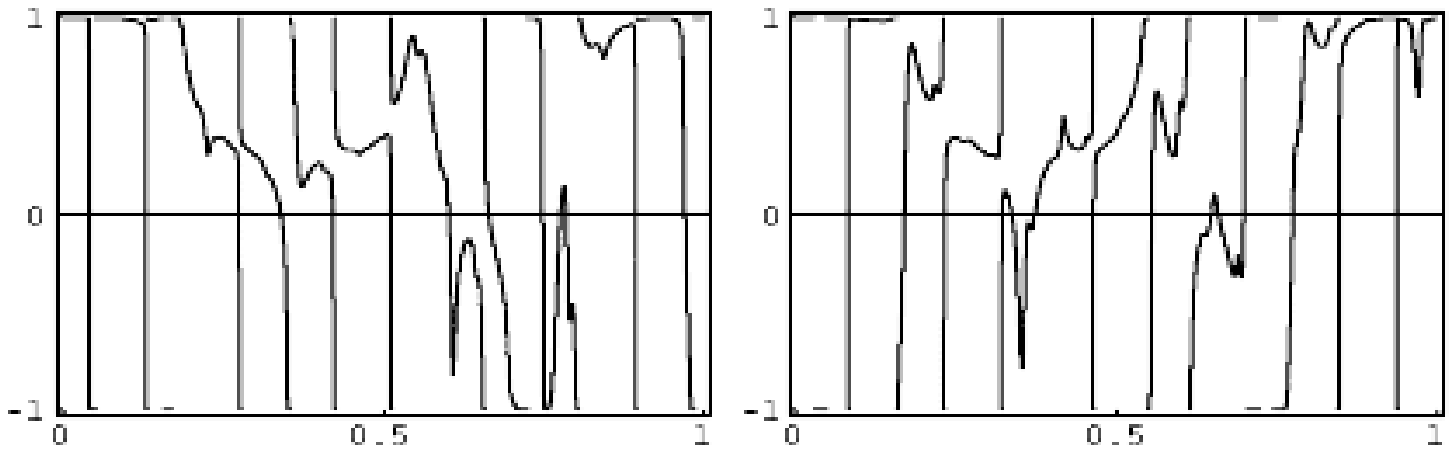}}
  \put(40,-1){$\phi$}
  \put(98,-1){$\phi$}
  \put(4,16){\rotatebox{90}{$\tanh(x)$}}
  \put(39,43){$(a)$}
  \put(98,43){$(b)$}
 \end{picture}
\caption{\footnotesize Maps for $\omega^*$ and $\epsilon=1$.
(a) $\eta=11$, $E=2.342$; (b)$\eta=10$, $E=2.347$.}
\label{fig:wind}
\end{figure}

\begin{figure}[ht]
 \begin{picture}(73,60)(-10,-10)
  \put(10,-15){\includegraphics[width=120mm]{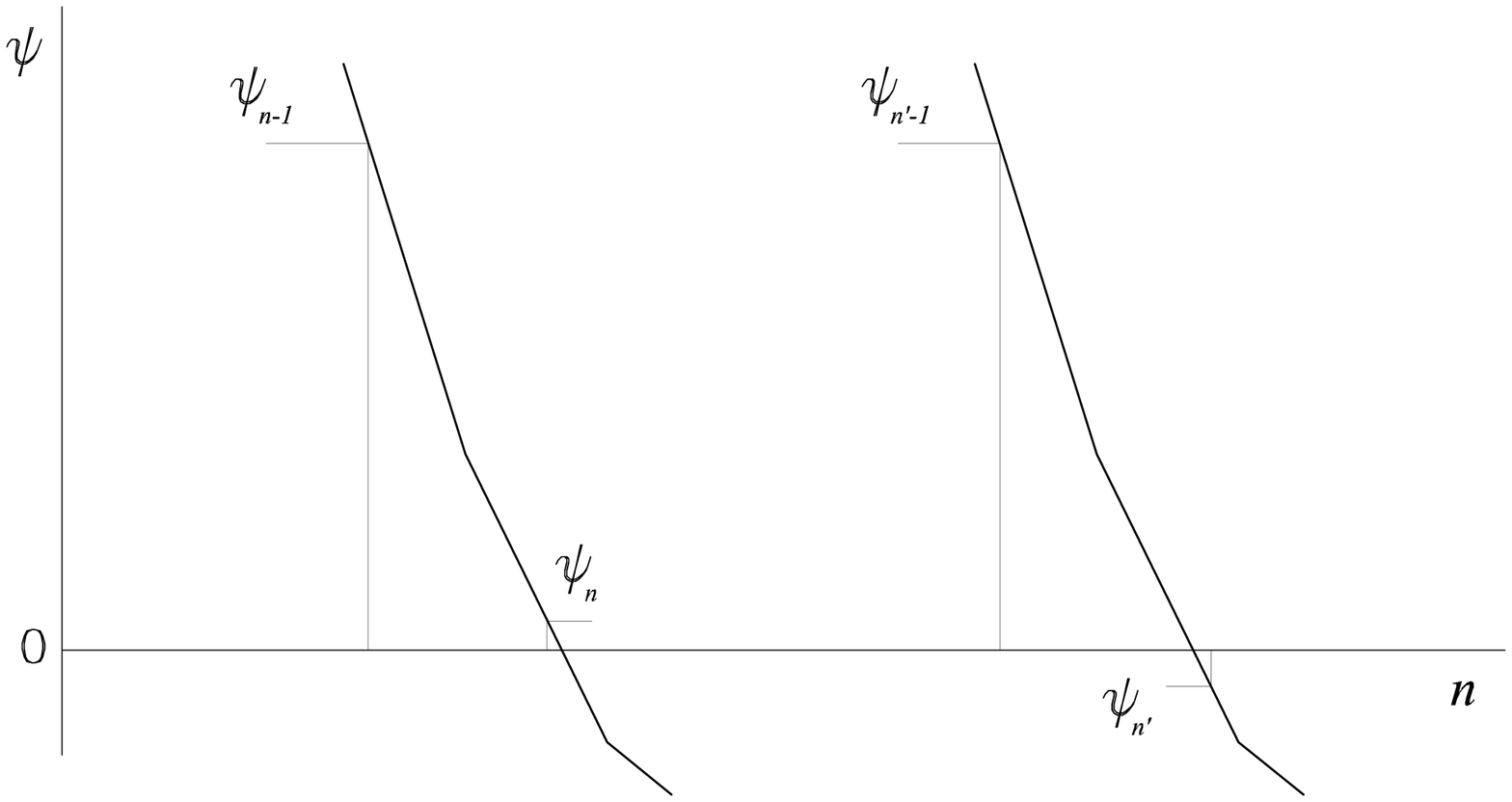}}
 \end{picture}
\caption{\footnotesize The wave function is null, $\psi=0$, at
two distinct spatial points but the corresponding $x$ have opposite signs.}
\label{fig:wave}
\end{figure}

The $x$ identification $x:\ +\infty\cong-\infty$ is somehow more subtle.
By inspecting the attractors we realize that the one-dimensional
attractors cross from $\pm\infty$ to $\mp\infty$ at the same value of $\phi$
(as an example see figures~\ref{fig:attractor} and~\ref{fig:wind}).
The $\phi\times x$ diagram gives us local properties of the original continuous variable $\psi$ at each spatial index $n$.
We note however that we completely loose all the information about the spatial localization itself
(ordering on the spatial indexes $n$) of where (and in which order) the original variable $\psi$ have those properties.
We are going to show that the $x_n=\psi_{n-1}/\psi_n$ variable is describing exactly
the same situation for both cases $x=+\infty$ and $x=-\infty$. The setup
here is the original continuous variable being null $\psi=0$ at some (or many) points,
$x_n=\psi_{n-1}/\psi_n=\pm\infty$ means that $\psi_n=0$.
To show the validity of this statement based on geometrical arguments see figure~\ref{fig:wave} where two
points, at different spatial localizations $n$ and $n'$, are considered,
the discretized points $\psi_{n-1}$, $\psi_{n}$, $\psi_{n'-1}$ and
$\psi_{n'}$ are represented in the figure. Take in both cases
$|\psi_{n-1}|\gg|\psi_n|$ and $|\psi_{n'-1}|\gg|\psi_n'|$.
This means that $x_n\sim+\infty$ and $x_n'\sim-\infty$ do correspond to the same
situation of the continuous variable $\psi=0$.
Both kind of situations will always occur for irrational
$\omega$ (many times for large $N$) because the
period of the variable $\psi$ and the period of the underlying
discretized lattice (given by the magnetic flux $\omega$) are \textit{incommensurate}~\cite{AZ_0,HO_0,PS_1}.

In order to formalize our argument we can invert the map $x_n=\psi_{n-1}/\psi_n$ between
variables obtaining $\psi_n=\psi_{n-1}/x_n$, then we have that $x_n\to \pm\infty$ correspond
to one only value $\psi_n\to 0$. Although not in the scope of the present study, it is
interesting to note that $x_n\to 0$ correspond to $\psi_n\to\pm\infty$, therefore a
divergence of the original variable $\psi$.

So we just proved that the variable identifications
$\phi:\ 0\cong 1$ and $x:\ +\infty\cong-\infty$ are natural with out need for any \textit{a priori}
assumptions. Then the space of variables $\phi\times x$ is topologically
a torus. Therefore the one-dimensional attractors, for irrational $\omega^*$,
are close contours living on the surface of that torus. By inspection we conclude that the
attractors wind several times along the $x$ coordinate.
Furthermore the winding number along the $x$ homology cycle is unique for each continuous Localized phase.
So we classify these several phases accordingly, such that our classification label $\eta$
is the winding number along the $x$ homology cycle of the one-dimensional attractors.
Note that the winding number in the $\phi$ homology cycle is always $1$
except for $E=0$ and $\epsilon<1$ where there are two one-dimensional attractors with
winding number $0$ in the $x$ axis and $1$ in the $\phi$ axis~\cite{PS_1}. This particular
case $E=0$ is excluded from our classification.

The great advantage of this classification is that it also provide us a natural hierarchy based on the areas
of the several phases. It coincides with the ordering of the areas occupied by the corresponding
Localized phases. Furthermore as we consider successive $\omega_m$'s, the order of appearance
of the phases also coincides with their hierarchical importance.
As explained in section~\ref{sec:phases} this is due to the Localized phases emerging from the splitting
of the Extended phases. The diagram for irrational $\omega^*$ and the classification
of areas up to $\eta=12$ is presented in figure~\ref{fig:phaseir}. As an example for higher
$\eta$ attractors we present in figure~\ref{fig:wind} the cases for $\eta=10$ and $\eta=11$.

\section{Hierarchical Fractal Structure}
Here we will define the generic rules to build the fractal structure
of the phase diagram in the first quadrant ($E>0$ and $\epsilon>0$).

First let us state some conclusions obtained by inspection (CI) of successive phase diagrams for approximants $\omega_m$
(see figures~\ref{fig:phaser} and~\ref{fig:phaseir}):
\begin{enumerate}
\item[CI.1] In each new diagram for successive approximants the
last phase on the far right (for $E>2$) has always $\eta=0$ and correspond to
the phase with greater area. It is always present and doesn't contribute anything to explain the structure.
So \textit{\bf $\mathbf{\eta=0}$ is excluded from the classification}.
\item[CI.2] The first phase to appear next (corresponding to the next greater area)
as we move in the Fibonacci succession, is $\eta=1$. Although we have two Fibonacci elements
corresponding to this number, we have only one phase in the phase diagram. In some
way (that we will explain later) this means that \textit{\bf $\mathbf{\eta=1}$ is a degenerate phase}.
\item[CI.3] We recall from section~\ref{sec:phases} that for each approximant $\omega_m$ there
are $F_{m-1}$ Localized phases, from the Fibonacci succession we have that
$F_m=F_{m-2}+F_{m-1}$, then as we go trough the approximants,
from $\omega_m$ to $\omega_{m+1}$ emerge exactly $F_{m-2}$ new Localized phases.
It seems then a good guess to organize the phases in levels such
that in each level $k=m-1$ there are exactly $F_{k-2}$ labels in the range
$F_{k-1}<\eta_{level\ k}\leq F_{k}$. We are therefore building a \textit{\bf triangular structure}.
\item[CI.4] Also as we increase $m$ the last phase on the far right (just before $\eta=0$)
corresponds always to one label belonging to the Fibonacci succession $\eta_k^R=F_m$
with $k=m-1$. We will call $\eta_k^R$ the \textit{\bf right pivots}.
\item[CI.5] The last phase on the left side of the diagram (near $E=0$)
only changes for the second odd number in the Fibonacci succession
(even, odd, \textit{odd}, even, odd, \textit{odd},\ldots). This means that
only for $k=3n+1$ (for all integer $n$) we will have a label $\eta_{3n+1}^L=F_{3n}+F_{3n-2}$.
We will call $\eta_k^L$ the \textit{\bf left pivots}.
\item[CI.6] As we move along the approximants new phases will emerge in the bulk of the diagram next to the ones that already exist
in the previous approximants. However if one phase $\eta_n$ appears at some approximant $\omega_m$, at the immediately next approximant $\omega_{m+1}$
no new phases will appear next to this phase $\eta_n$. So, for each new phase appearing, \textit{\bf there are not adjacent new phases
at the immediately next approximant}.
\item[CI.7] Also these new phases appearing adjacent to the already existing phases in the bulk will appear either on the right or on the
left of the already phases that already exist. So the \textit{\bf order of appearance of new phases alternate between left and right}.
\end{enumerate}

So we have a triangular Hierarchical Tree organized in levels, each \textit{level~k} has exactelly $F_{k-2}$ elements
(see point~CI.3 above) and we are ready to define the rules to build it.
In order to accomplish it we first give the rules for the edges of the triangle and then
define a recursive rule for the bulk of the triangle.
For the reader interested in the full details we present in appendix~\ref{A.A}
the Lyapunov exponents for $\omega=\omega^*$ and $\epsilon=1$ with enough resolution
to identify the several phases (that correspond to each negative \textit{bump}) and
in appendix~{\ref{A.B}} we list the $\phi\times x$ diagrams up to $\eta=55$
with indication of the respective energies.

We can define the vertex and edges of the triangle following the following rules:
\begin{enumerate}
\item[R.I] On the top vertex of the triangle is placed the degenerate label $\eta=1$ (see point~CI.2 in our list of conclusions above).
It is degenerate simply because it belongs simultaneously to the right and left edge of the triangle, i.e. it is both a
left and a right pivot $\eta^R_0=\eta^L_0=1$ (see conclusions by inspection~CI.2,~CI.4 and~CI.5).
\item[R.II] At the right edge of the triangle we have at each \textit{level~k} the right pivots $\eta^R_{k}=F_{k+1}$ distributed, in sequence
(see point~CI.4 on the list of conclusions above).
\item[R.III] On the left edge not all levels have a pivot, we will have left pivots only for \textit{level~(3n+1)} (for any integer $n$)
with labels $\eta^L_{3n+1}=F_{3n}+F_{3n-2}$ (see figure~\ref{fig:hier}).
\end{enumerate}

Next we will define a recursive rule for the bulk of Tree. In order to distribute
the labels in each level, consider the Tree completely built up to \textit{level~k}
and let us construct the next \textit{level~k+1}. We have therefore $F_{k-1}$ labels
in the range $F_{k}<\eta_{level~k+1}\leq F_{k+1}$ (see point~CI.3). Although we already have rules for the
edges of the triangle (see Rule~II and Rule~III) we include in the recursive rules also the edges. The following rules
should be applied in sequence:
\begin{enumerate}
\item[R.1] Group the two last right pivots $(\eta^R_{k-1},\eta^R_k)$ and sum them together,
the result is the new pivot of level $k+1$, i.e. $\eta^R_{k+1}=\eta^R_{k-1},\eta^R_k$.
\item[R.2] Group in pairs all the remaining elements (the labels) up to level $k-1$ (we note that no other elements of \textit{level~k} will be
used except for the right pivot $\eta^R_k$). The way to group them is to start from the right
pivot $\eta^R_{k-2}$ and group it with the immediately adjacent element to the left. Then we take the next element immediately adjacent to
the left and group it to next one adjacent to the left, and so on. Here adjacent means the order in the phase diagram as we move to the left,
in terms of the Hierarchical Tree, i.e. the closest element in relation to the horizontal projected distance. From two in two levels
there will be out of the pairs a single left pivot (see point~CI.5 above), then after applying Rule~R.3 below, apply also Rule~R.4.
\item[R.3] Sum the right pivot of \textit{level~k} ($\eta^R_k$) to all the pairs obtained in Rule~2, then we swap the order
of the elements inside each pair and place the resulting (swapped elements) pair at level $k+1$ between the elements of the original pair.
Again, here, between means in relation to the horizontal projection of the elements. We note that these rules do reproduce
the behaviour described in the conclusions by inspection~CI.6 and~CI.7.
\item[R.4] From two in two levels we will have a single left pivot unpaired of \textit{level~k-1},
in this case we sum it to the right pivot $\eta^R_k$ and place the result as the left pivot of \textit{level~k+1}.
\end{enumerate}

We note that in order to apply the recursive rules we only need the two first labels, i.e. $\eta^R_0=\eta^L_0=1$ and $\eta^R_1=2$.
We note that Rule~II is equivalent to Rule~1 and Rule~III to Rule~4, we simple list both set for completeness.
In this way the rules defining the triangular structure
are complete! The first 9 levels of the Hierarchical Tree are shown in figure~\ref{fig:hier}, we can compare it with
the corresponding phases in the phase diagram (figure~\ref{fig:phaseir}).
In appendix~\ref{A.C} we show graphically how to build the triangular structure up to level 9 using these
recursive rules.
\begin{figure}[ht]
 \begin{picture}(100,190)(25,7)
  \put(0,2){\rotatebox{90}{\includegraphics[width=190mm,height=180mm]{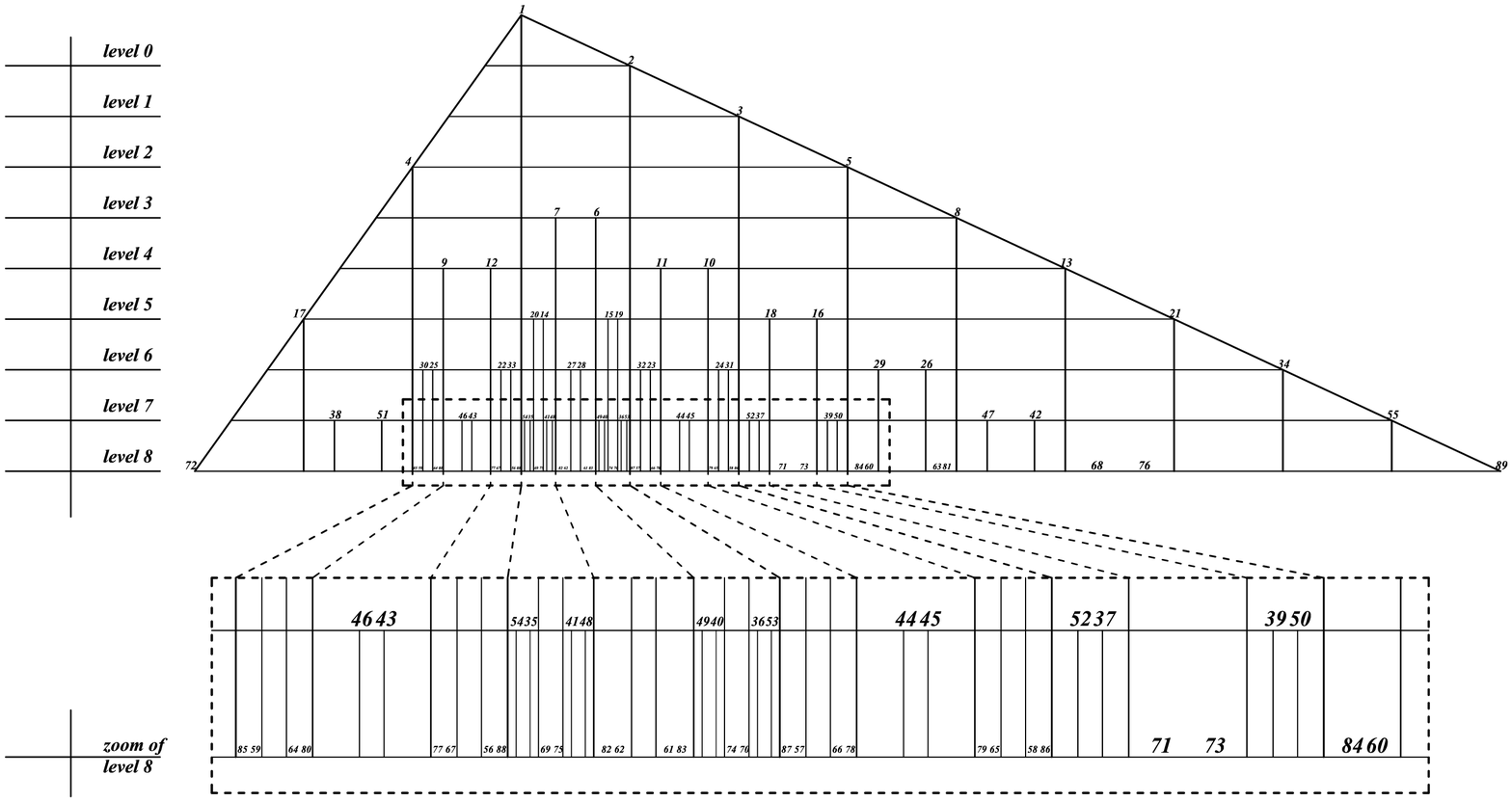}}}
 \end{picture}
\caption{Fractal structure of the phase diagram (to be seen in landscape orientation).}
\lb{fig:hier}
\end{figure}

\section{Conclusions}
In this paper we have proposed an integer
topological hierarchical classification for the infinite many Localized phases in the phase diagram $E\times\epsilon$
of the Harper map for irrational parameter $\omega^*=(\sqrt{5}-1)/2$. This map is equivalent to the Harper equation which,
among other applications, describes the Azbel-Hofstader model with $E$ being the energy, $\epsilon$ the anisotropy parameter
and $\omega$ the magnetic flux per unit cell.
The classification is based in the winding number around the holonomy cycles of the
One-dimensional attractors in the toroidal space of variables $\phi\times x$.
For the classifications obtained, the fractal structure of the 
$E\times\epsilon$ phase diagram is built and organized into a
triangular structure that exactly describes the phase diagram taking
in account the ordering of the several phases. Moreover the
integer classification labels (1, 2, 3, \ldots) exactly correspond
to the phase importance, concerning their area in the phase diagram and order of appearance 
in the succession of phase diagrams for the several approximants $\omega_n$ converging to $\omega^*$.
So in this since the triangular fractal structure correctly describes all approximants. 
As possible future works in this subject it would be interesting to investigate the fractal structure for $\omega$ given
by other irrational numbers and, if possible, to find a rule for generic irrational numbers and respective approximants.

\clearpage
\textbf{Acknowledgements}\\
This work was supported by PRAXIS XXI/BD/11461/97 and SFRH/BPD/17683/2004.
The first part of this work was done at the Department of Physics of the University of Oxford, without the
well organized computer resources efficiently available there it would be impossible to accomplish it.
The author thanks F.P. Mancini and M.H.R. Tragtenberg for several discussions and comments.

\newpage
\appendix

\section{Lyapunov Exponent\lb{A.A}}
In this appendix we present the Lyapunov exponents for $\epsilon=1$ and $\omega^*$. The
several phases are labelled until $\eta=55$.
\begin{figure}[ht]
 \begin{picture}(130,150)(-4,5)
  \put(-8,0){\includegraphics[width=170mm]{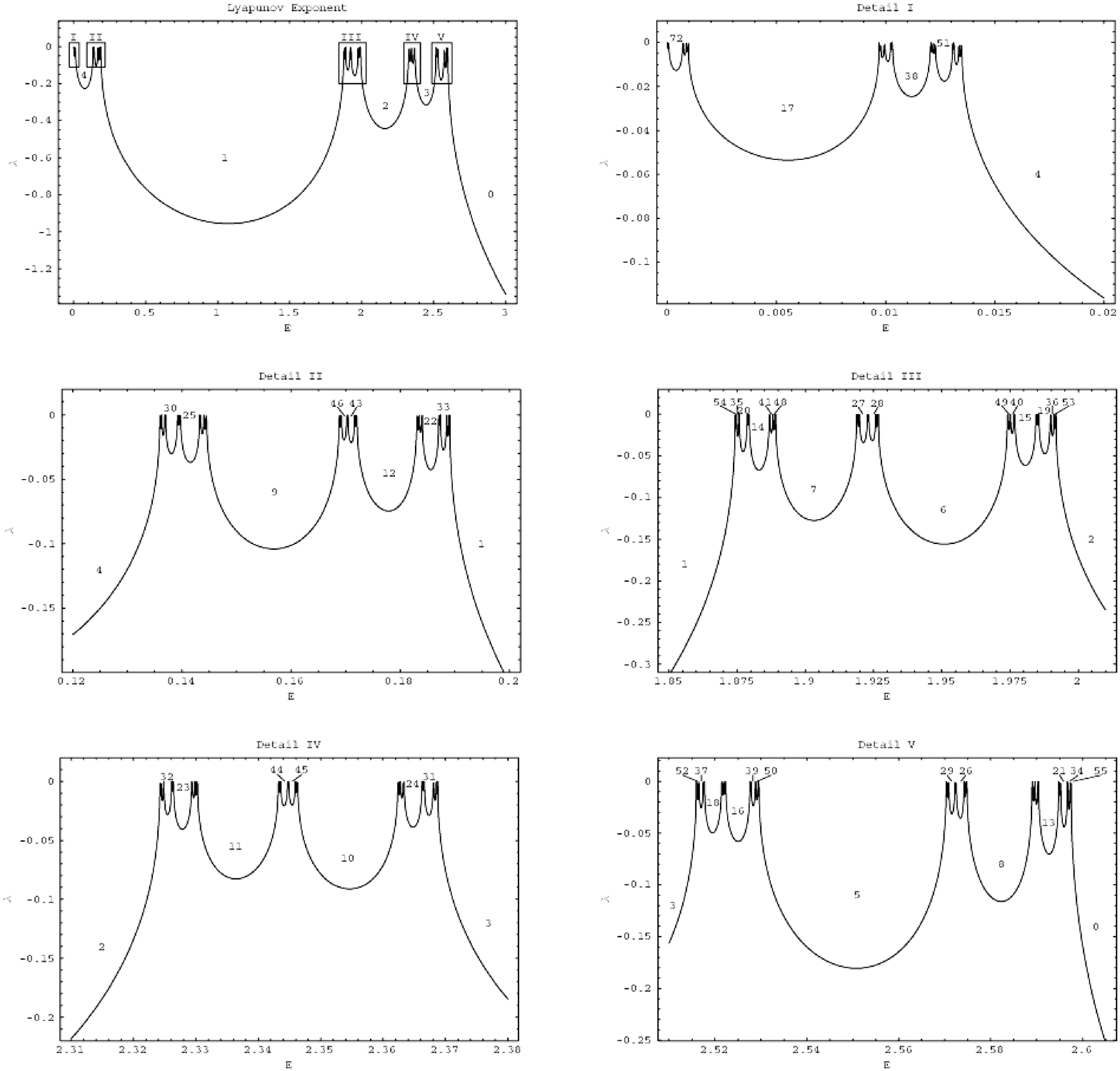}}
 \end{picture}
\caption{Lyapunov exponent $\lambda$ as a function of energy $E$ for $\epsilon=1$ and $\omega=\omega^*$.
The Details I to V correspond to the marked regions in the first graph.}
\label{A.lya}
\end{figure}

\section{$\phi\times x$ diagrams\lb{A.B}}
Next we list all the attractors in the space of variables
$\phi\times x$ up to $\eta=55$.

\begin{figure}[ht]
 \begin{picture}(150,210)(-4,10)
  \put(0,2){\includegraphics[width=155mm]{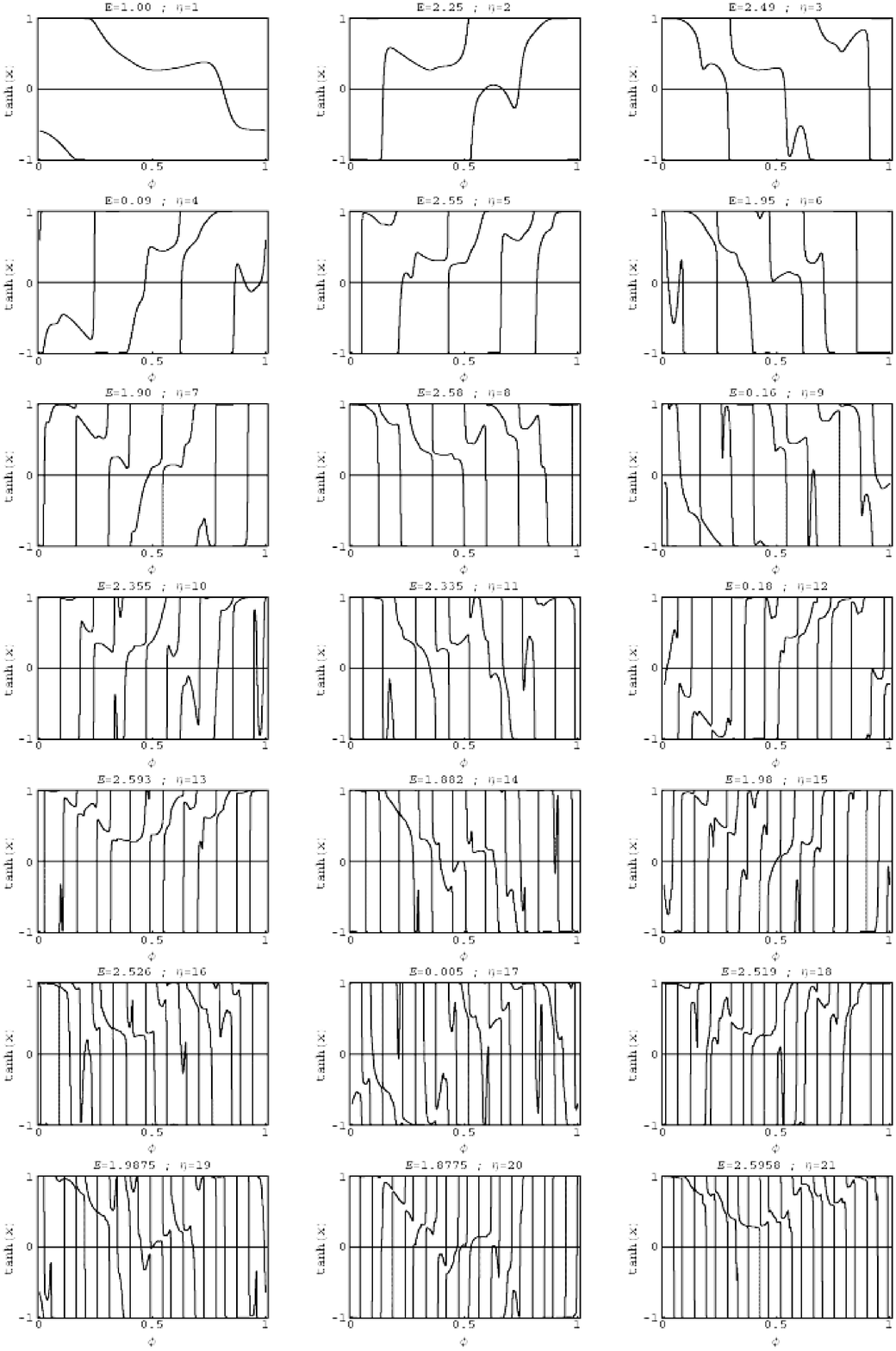}}
 \end{picture}
\caption{Attractors in the space of variables $\phi\times x$ up to $\eta=21$.}
\end{figure}

\begin{figure}[ht]
 \begin{picture}(150,210)(-4,10)
  \put(0,2){\includegraphics[width=155mm]{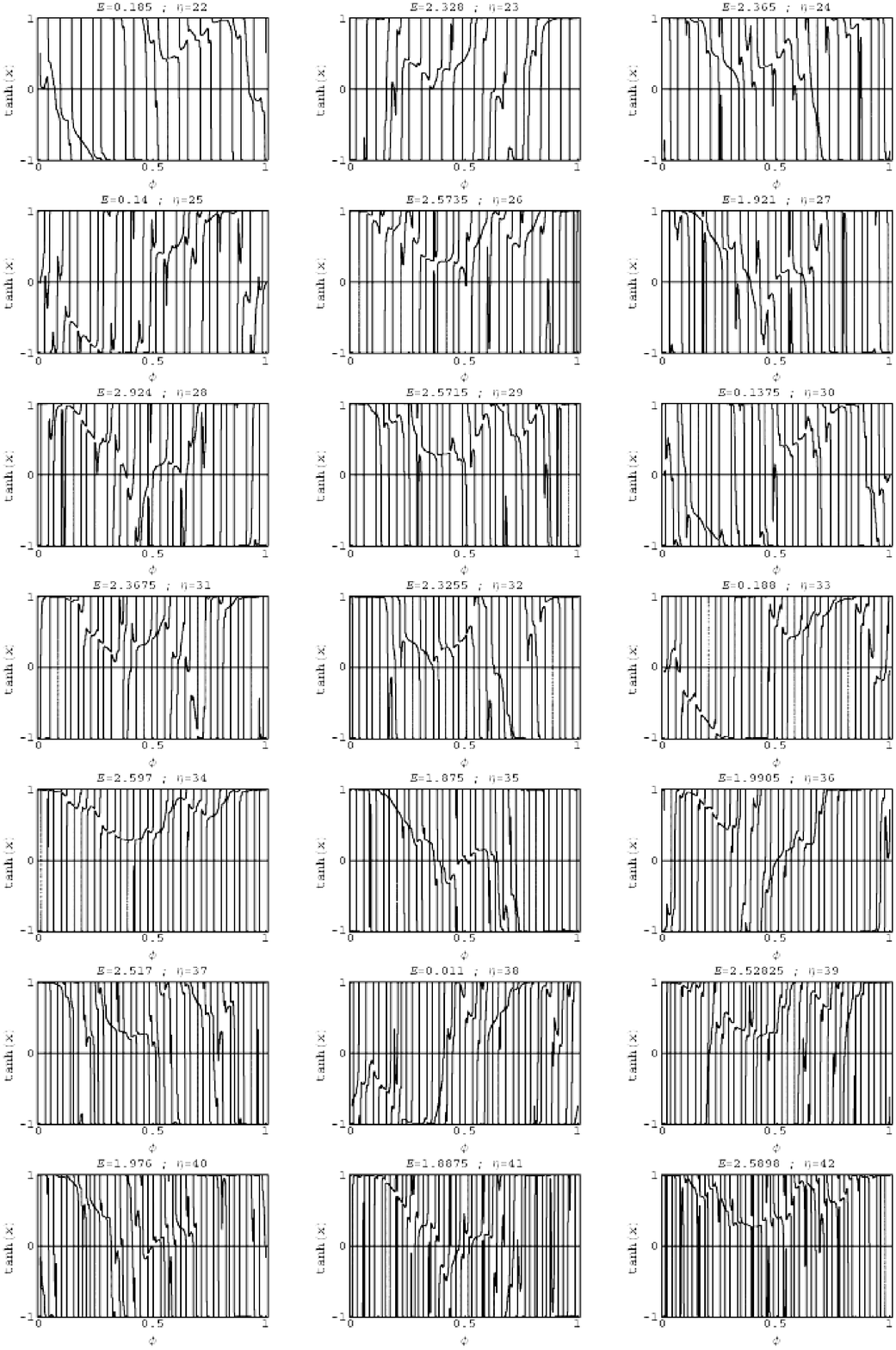}}
 \end{picture}
\caption{Attractors in the space of variables $\phi\times x$ from $\eta=22$ to $\eta=42$.}
\end{figure}

\begin{figure}[ht]
 \begin{picture}(150,210)(-26,10)
  \put(0,2){\includegraphics[width=105mm]{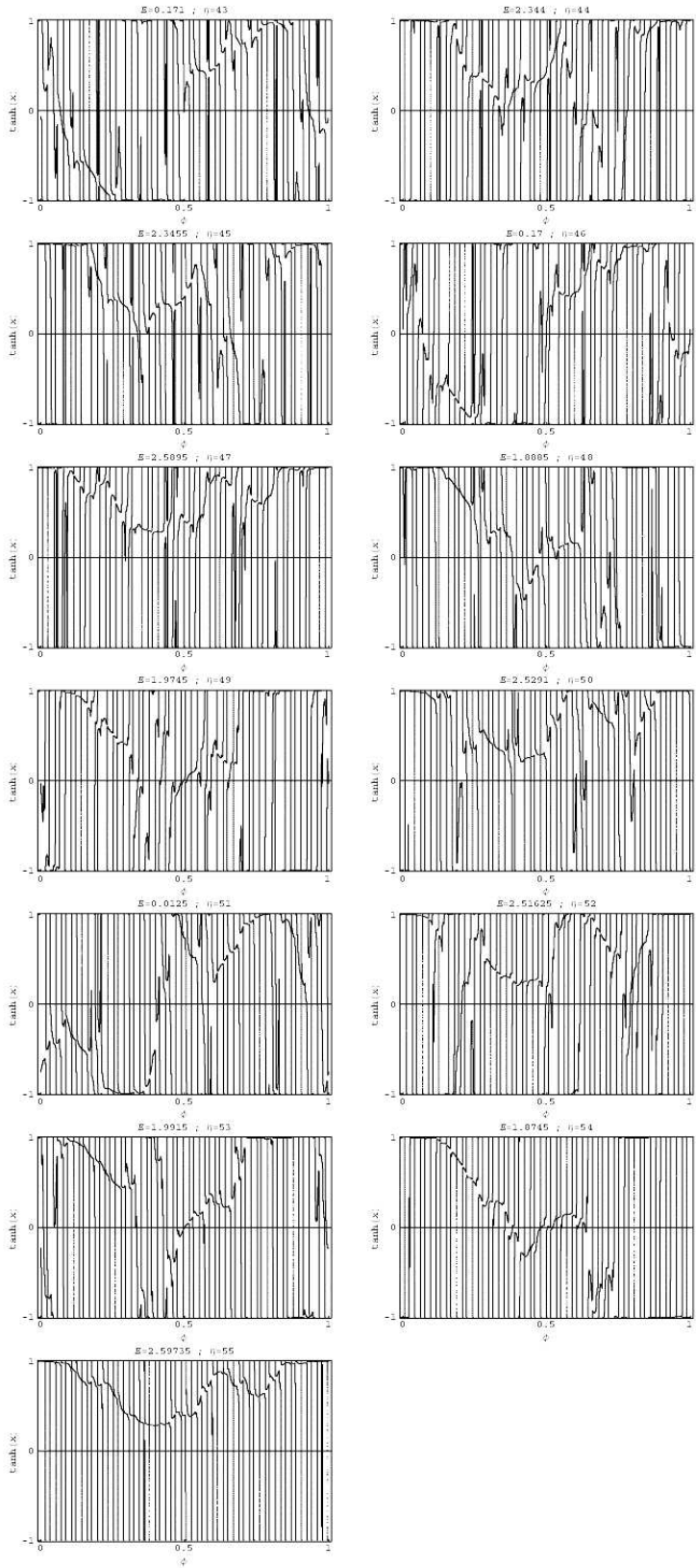}}
 \end{picture}
\caption{Attractors in the space of variables $\phi\times x$ from $\eta=43$ to $\eta=55$.}
\end{figure}

\clearpage

\section{Building the Hierarchical Tree\lb{A.C}}
We show graphically how to build the Hierarchical Tree up to \textit{level~9}.
\textit{Level~1} and \textit{Level~2} are given and only contain the vertex (simultaneously left and right) pivot
$\eta^R_0=\eta^L_0=F_1=1$ and the right pivot $\eta^R_1=F_2=2$. We give details of the application of the
construction rules up to \textit{level~6} which contains all rules.
The remaining levels do not involve any new rule and can be build straight forwardly.

On the following figures the pivot used to construct each level $n+1$ (i.e. $\eta^R_n$, the pivot of the level
immediately above the one we are constructing) is inside a darker circle. The remaining elements used are grouped
in pairs pictorially marked. We note that up to some \textit{level~n} we have present in our Tree all labels up to $\eta^R_n=F_{n+1}$.

\begin{figure}[ht]
 \begin{picture}(150,40)(-4,45)
  \put(0,2){\includegraphics[width=155mm]{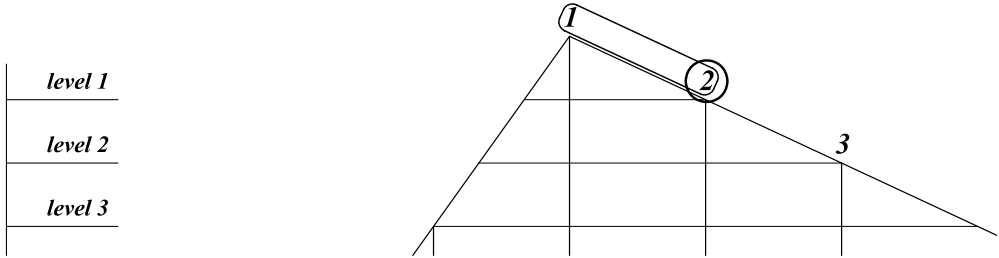}}
 \end{picture}
\caption{Building \textit{level~2} ($n+1=2$). The right pivot at level $n=1$ is $\eta^R_1=F_2=2$.
The next pivot is obtained by adding the two previous pivots $\eta^*_2=\eta^R_0+\eta^R_1=F_1+F_2=1+2=3$.
There are no more elements in this level.}
\end{figure}

\begin{figure}[ht]
 \begin{picture}(150,40)(-10,30)
  \put(0,2){\includegraphics[width=135mm]{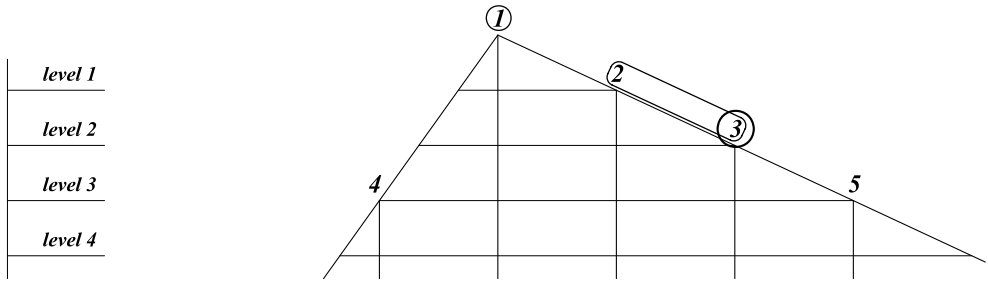}}
 \end{picture}
\caption{Building \textit{level~3} ($n+1=3$). The pivot at level $n=2$ is $\eta^R_2=F_3=3$.
The next pivot is obtained by adding the two previous pivots $\eta^R_3=\eta^*_1+\eta^*_2=2+3=5$.
There are no enough elements to group in pairs, we have one single element at the vertex of the triangle $\eta^L_0=\eta^R_0=F_1=1$
that we deal as being a left pivot. Then we sum it the right pivot of \textit{level~2} $\eta^R_2$ obtaining the new
left pivot $\eta^L_3=\eta^L_0+\eta^R_2=F_1+F_3=1+3=4$.}
\end{figure}

\begin{figure}[ht]
 \begin{picture}(150,40)(-4,35)
  \put(0,2){\includegraphics[width=140mm]{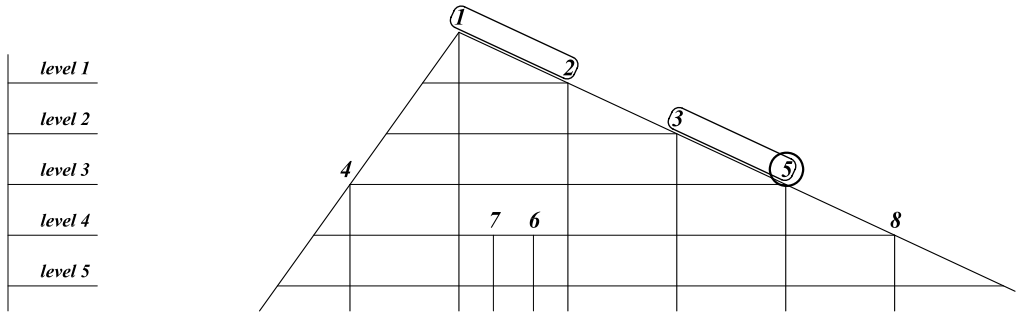}}
 \end{picture}
\caption{Building \textit{level~4} ($n+1=4$). The pivot at level $n=3$ is $\eta^R_3=F_4=5$.
The next pivot is obtained by adding the two previous pivots $\eta^R_4=\eta^*_2+\eta^*_3=3+5=8$.
The remaining elements on the levels above (not including) \textit{level~3} are grouped in pairs, we obtain
one single pair $(1,2)$. We sum this pair to the current level pivot $\eta^R_3=5$ and place the resulting pair
with the order of the elements reversed below and between the elements of the original pair $(1,2)$.
This means we obtain the pair $(7,6)$ and place it as shown in the figure.}
\end{figure}
\begin{figure}[ht]
 \begin{picture}(150,40)(-8,25)
  \put(0,2){\includegraphics[width=140mm]{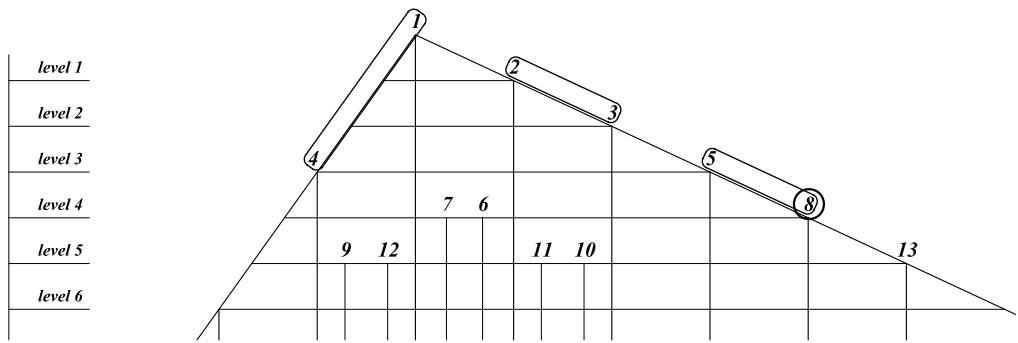}}
 \end{picture}
\caption{Building \textit{level~5} ($n+1=5$). The pivot at level $n=4$ is $\eta^R_4=F_5=8$.
The next pivot is obtained by adding the two previous pivots $\eta^R_5=\eta^*_3+\eta^*_4=5+8=13$.
The remaining elements on the levels above (not including) \textit{level~4} are grouped in pairs, we obtain
two pairs $(4,1)$ and $(2,3)$. We sum these pairs to the current level pivot $\eta^R_4=8$ and place the resulting pairs
with the order of the elements reversed below and between the elements of the original pairs.
This means we obtain the pairs $(9,12)$ and $(11,10)$ and place them as shown in the figure.}
\end{figure}
\begin{figure}[ht]
 \begin{picture}(100,180)(12,5)
  \put(0,2){\rotatebox{90}{\includegraphics[width=190mm]{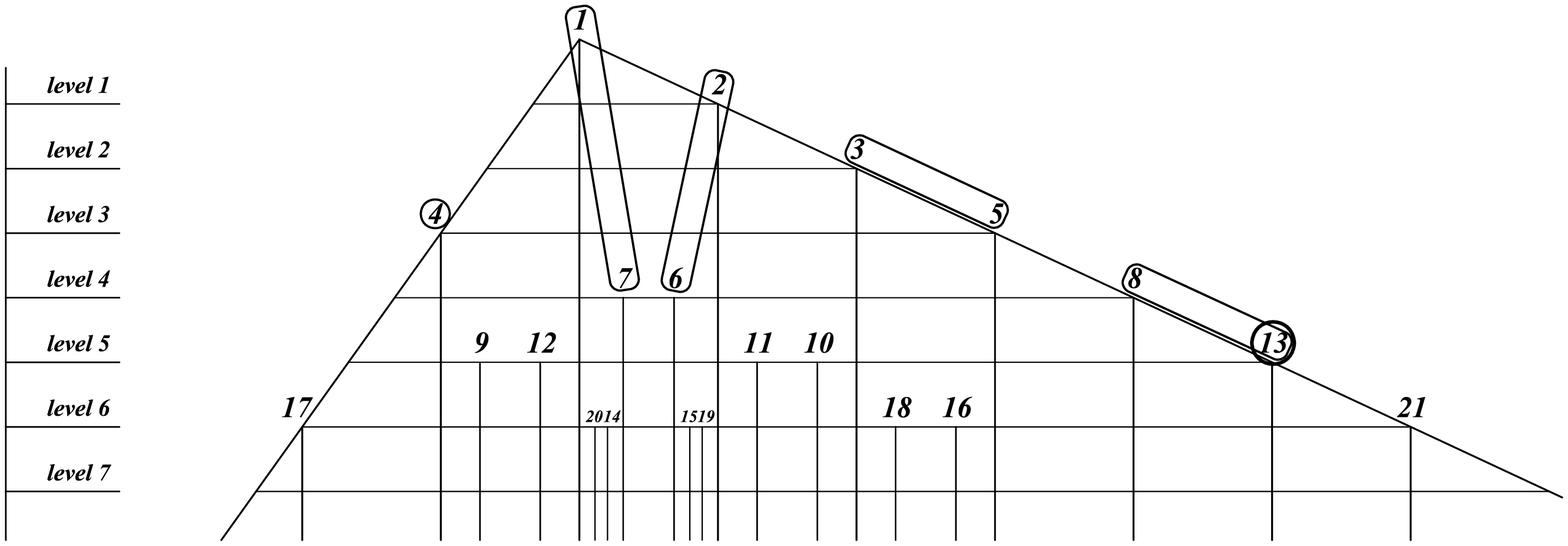}}}
 \end{picture}
\caption{Building \textit{level~6} ($n+1=6$). The pivot at level $n=5$ is $\eta^R_5=F_6=13$.
The next pivot is obtained by adding the two previous pivots $\eta^R_5=\eta^*_3+\eta^*_4=8+13=21$.
The remaining elements on the levels above (not including) \textit{level~5} are grouped in pairs from right to left.
As expected for \textit{level~6} we have one lonely left pivot $\eta^L_3=4$ and we obtain three pairs $(1,7)$, $(6,2)$ and $(3,5)$.
We sum the current level pivot $\eta^R_5=13$ to the lonely left pivot obtaining the new left pivot $\eta^L_6=\eta^R_5+\eta^L_3=13+4=17$.
As for the three pairs we sum them to the current level pivot $\eta^R_5=13$ and place the resulting pairs
with the order of the elements reversed below and between the elements of the original pairs.
This means we obtain the pairs $(20,14)$, $(15,19)$ and $(18,16)$ and place them as shown in the figure
(to be seen in landscape orientation).}
\end{figure}

\begin{figure}[ht]
 \begin{picture}(100,210)(20,8)
  \put(0,2){\rotatebox{90}{\includegraphics[height=174mm]{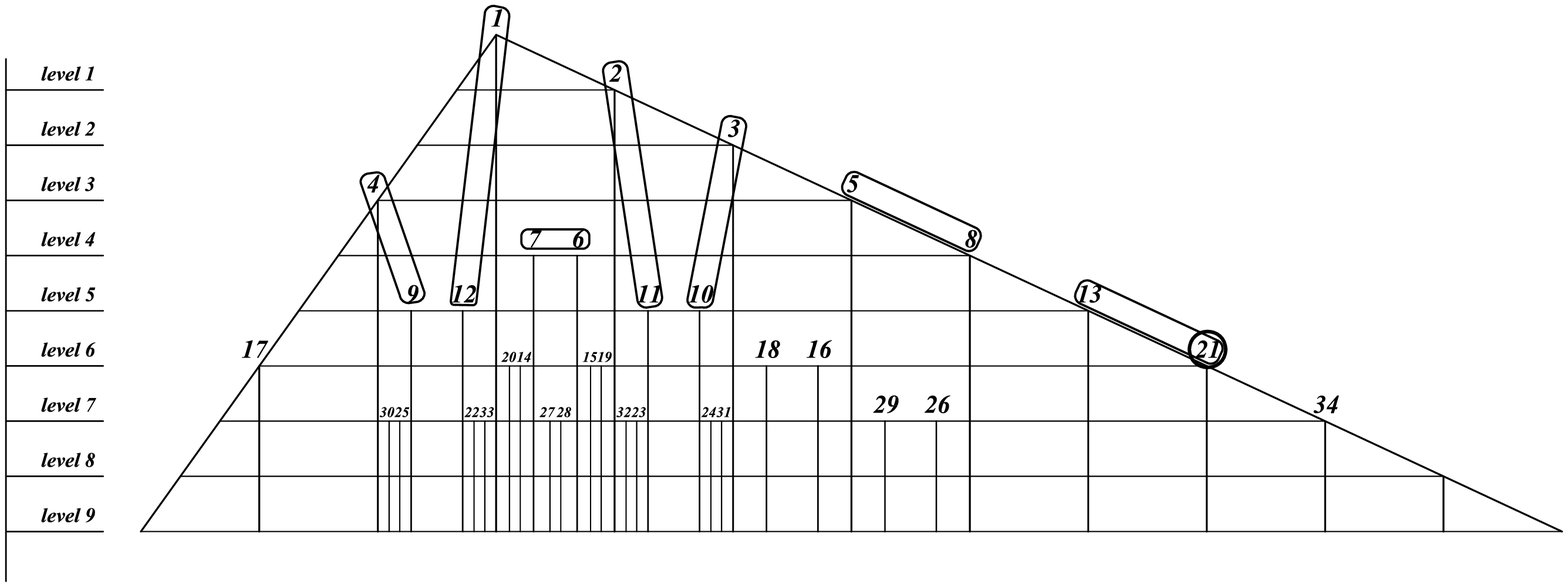}}}
 \end{picture}
\caption{Building \textit{level~7} (to be seen in landscape orientation).}
\end{figure}
\begin{figure}[ht]
 \begin{picture}(100,210)(20,8)
  \put(0,2){\rotatebox{90}{\includegraphics[height=174mm]{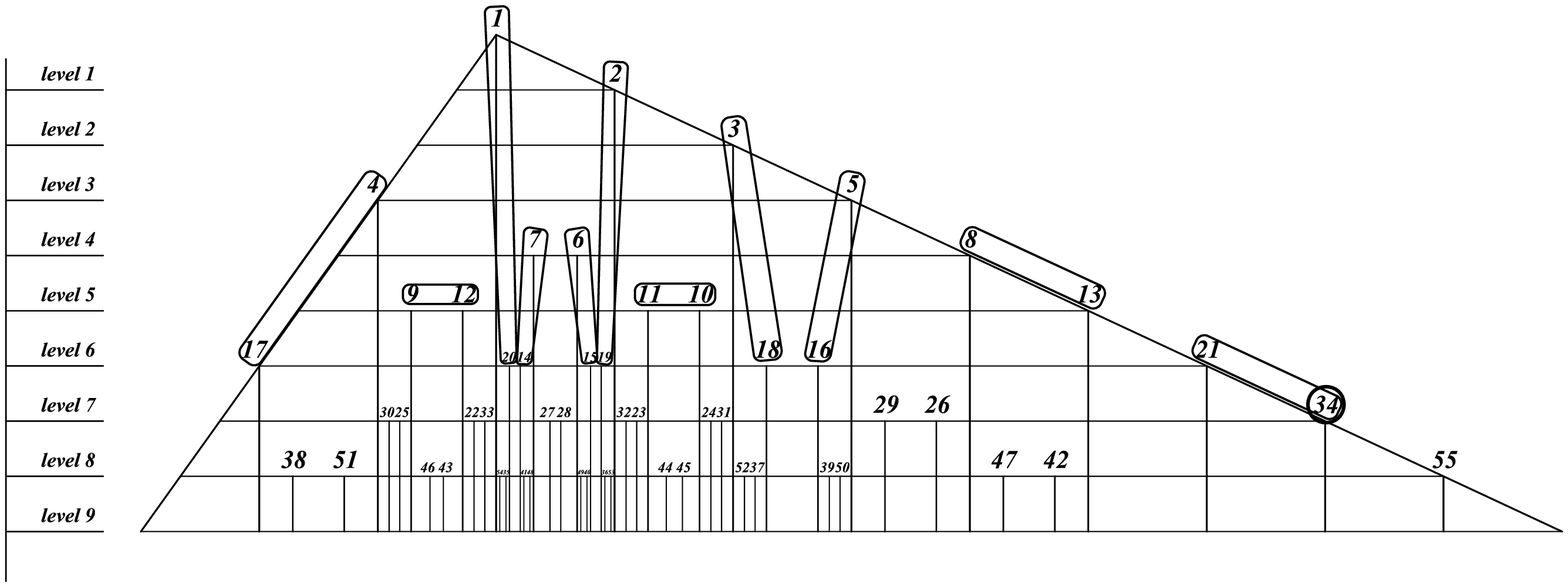}}}
 \end{picture}
\caption{Building \textit{level~8} (to be seen in landscape orientation).}
\end{figure}
\begin{figure}[ht]
 \begin{picture}(100,210)(33,8)
  \put(10,2){\rotatebox{90}{\includegraphics[height=174mm]{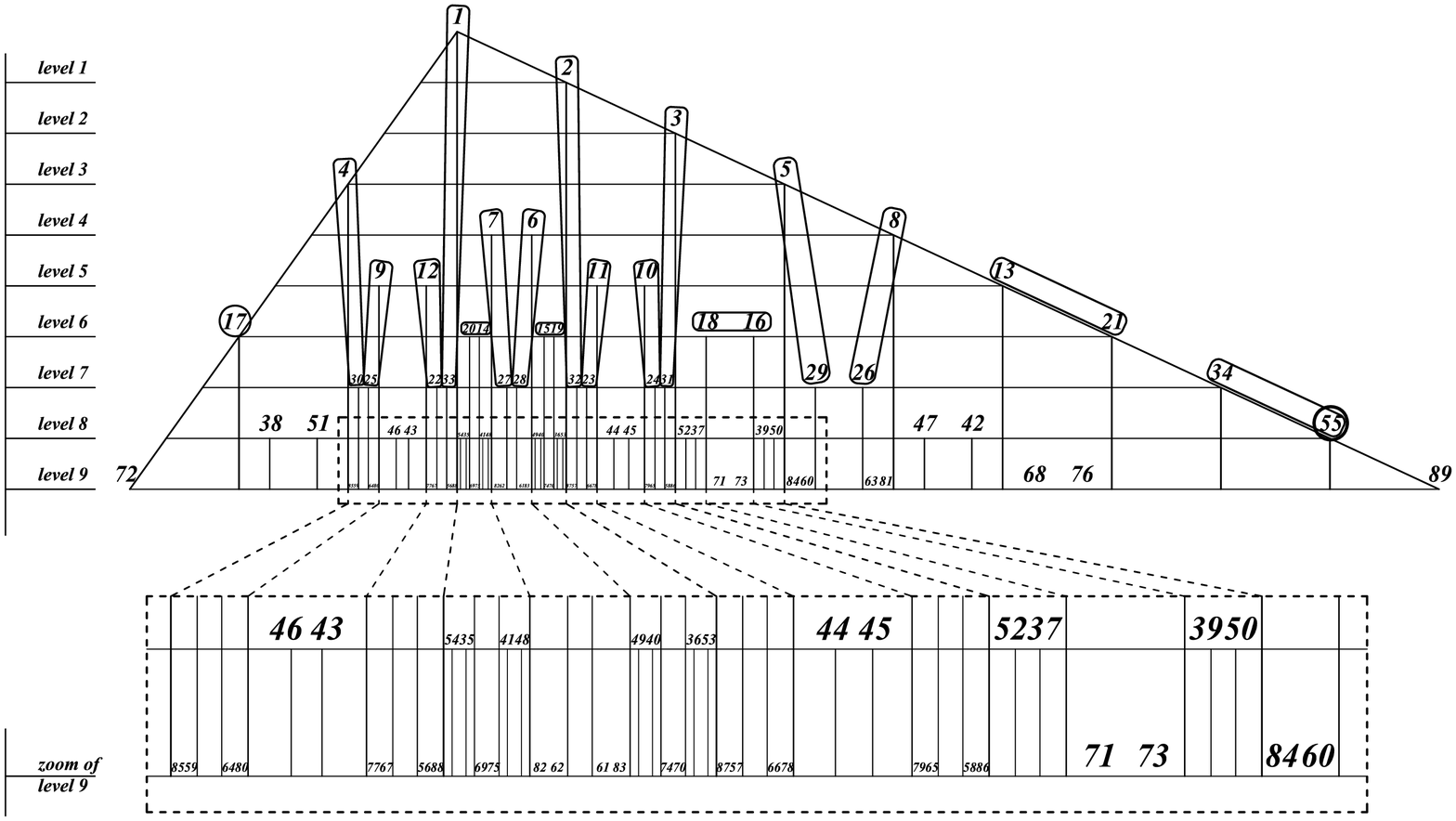}}}
 \end{picture}
\caption{Building \textit{level~9} (to be seen in landscape orientation).}
\end{figure}

\end{document}